\documentclass[preprints,article,accept,pdftex,moreauthors]{Definitions/mdpi}

\firstpage{1} 
\pubvolume{1}
\issuenum{1}
\articlenumber{0}
\pubyear{2025}
\copyrightyear{2025}
\datereceived{2 November 2025} 
\daterevised{28 November 2025} 
\dateaccepted{10 December 2025}
\datepublished{ } 
\hreflink{https://doi.org/} 

\usepackage{isotope}
\usepackage{braket}
\usepackage{siunitx}
\usepackage{graphicx}
\DeclareSIUnit\torr{Torr} 
\DeclareSIUnit\gauss{G}

\Title{\textls[-25]{Single Sr Atoms in Optical Tweezer Arrays for Quantum Simulation}}

\TitleCitation{Single Sr Atoms in Optical Tweezer Arrays for Quantum Simulation}

\Author{Veronica Giardini $^{1,2,\dagger}$\orcidA{}, Luca Guariento $^{3,4,\dagger,\ddagger}$\orcidB{}, Andrea Fantini $^{2,\dagger}$\orcidC{}, Shawn Storm $^{2}$\orcidD{}, Massimo Inguscio $^{1,3}$\orcidE{}, Jacopo Catani $^{3,1}$\orcidF{}, Giacomo Cappellini $^{3,1,\ddagger}$\orcidG{}, Vladislav Gavryusev $^{2,1,}$*\orcidH{} and Leonardo Fallani $^{2,1,3}$\orcidI{}}

\AuthorNames{Veronica Giardini, Luca Guariento, Andrea Fantini, Shawn Storm, Massimo Inguscio, Jacopo Catani, Giacomo Cappellini, Vladislav Gavryusev and Leonardo Fallani}

\isAPAStyle{%
       \AuthorCitation{Lastname, F., Lastname, F., \& Lastname, F.}
         }{%
        \isChicagoStyle{%
        \AuthorCitation{Lastname, Firstname, Firstname Lastname, and Firstname Lastname.}
        }{
        \AuthorCitation{Giardini, V.; Guariento, L.; Fantini, A.; Storm, S.; Inguscio, M.; Catani, J.; Cappellini, G.; Gavryusev, V.; Fallani, L.}
        }
}

\address{%
$^{1}$ \quad European Laboratory for Non-Linear Spectroscopy (LENS), University of Florence, via N. Carrara 1, \mbox{50019 Sesto Fiorentino, Italy;}\\
$^{2}$ \quad Department of Physics and Astronomy, University of Florence, via G. Sansone 1, \mbox{50019 Sesto Fiorentino, Italy;}\\
$^{3}$ \quad National Institute of Optics (CNR-INO), National Research Council, via N. Carrara 1, \mbox{50019 Sesto Fiorentino, Italy;}\\
$^{4}$ \quad Department of Physics Ettore Pancini, University of Napoli Federico II, via Cinthia 21, \mbox{80126 Napoli, Italy}}

\corres{Correspondence: vladislav.gavryusev@unifi.it}

\firstnote{These authors contributed equally to this work.}
\secondnote{Current address: Eniquantic S.p.A., via Ostiense 72, 00154 Roma, Italy}  

\abstract{
We report on the realization of a platform for trapping and manipulating individual \isotope[88]{Sr} atoms in optical tweezers. 
A first cooling stage based on a blue shielded magneto-optical trap (MOT) operating on the $\ket{^1\mathrm{S}_0} \rightarrow \ket{^1\mathrm{P}_1}$ transition at $\qty{461}{\nano\meter}$ enables us to trap approximately $\qty{4e6}{}$ atoms at a temperature of $\qty{6.8}{\milli\kelvin}$. Further cooling is achieved in a narrow-line red MOT using the $\ket{^1\mathrm{S}_0} \rightarrow \ket{^3\mathrm{P}_1}$ intercombination transition at $\qty{689}{\nano\meter}$, bringing $\qty{5e5}{}$ atoms down to $\qty{5}{\micro\kelvin}$ and reaching a density of $\qty{4e10}{\per\centi\meter\cubed}$. 
Atoms are then loaded into $\qty{813}{\nano\meter}$ tweezer arrays generated by crossed acousto-optic deflectors and tightly focused onto the atoms with a high-numerical-aperture objective.
Through light-assisted collision processes we achieve the collisional blockade, which leads to single-atom occupancy with a probability of about $50\%$. The trapped atoms are detected via fluorescence imaging with a fidelity of $99.986(6)\%$, while maintaining a survival probability of $97(2)\%$. The release-and-recapture measurement provides a temperature of $\qty{12.92(5)}{\micro\kelvin}$ for the atoms in the tweezers, and the ultra-high-vacuum environment ensures a vacuum lifetime higher than 7 min. These results demonstrate a robust alkaline-earth tweezer platform that combines efficient loading, cooling, and high-fidelity detection, providing the essential building blocks for scalable quantum simulation and quantum information processing with Sr atoms.
}



\begin{document}

\section{Introduction}
\label{sec:intro}

Arrays of single neutral atoms trapped in optical tweezers have emerged as a versatile platform for quantum science, enabling advances in quantum simulation, quantum computation, and metrology~\citep{Zhang2016a, Bernien2017, Browaeys2020, YagoMalo2024, SalesRodriguez2025, Chiu2025, Zhou2025, Wolswijk2025}. These systems offer a unique combination of precise atomic control and scalability, making them ideally suited to investigate many-body quantum phenomena. Recent years have witnessed rapid progress, including demonstrations of quantum simulation with Rydberg interactions~\citep{Saffman2010, Levine2019, Scholl2022, Bornet2023} and entangling operations between individually trapped atoms~\citep{Madjarov2020, Graham2022a}.
Alkaline-earth(-like) atoms, such as Sr and Yb, offer unique advantages for quantum science. Their narrow optical transitions enable efficient laser cooling and form the basis of state-of-the-art optical atomic clocks~\citep{Bloom2014, Finkelstein2024}. The presence of long-lived metastable states grants access to rich electronic structures and enables advanced quantum state control~\citep{Ye2008, Fukuhara2007, Scholl2025}. These properties make alkaline-earth(-like) atoms ideal for realizing quantum simulation, precision metrology, and quantum information platforms~\citep{Daley2011, Norcia2018, Nicholson2015}. Recent developments have extended these systems to optical tweezer arrays~\citep{Cooper2018, Saskin2019, Wu2022}, where narrow-line cooling, long coherence times, and strong Rydberg interactions allow for single-atom resolved control and imaging~\citep{Madjarov2020, Wilson2022, Gavryusev2016, Gavryusev2016a, FerreiraCao2020, Morandi2025}. These developments establish alkaline-earth tweezer arrays as a promising setting to explore strongly correlated quantum matter, realize programmable entangling operations, and interface many-body systems with precision metrology.

Here, we present an optical tweezer platform based on individually trapped bosonic strontium (\isotope[88]{Sr}) atoms. First, cold ensembles prepared via sequential blue and red magneto-optical traps (MOT) are loaded into arrays of tightly focused $\qty{813}{\nano\meter}$ tweezers. Then, light-assisted collisions ensure collisional blockade, leading to single-atom occupancy with a probability of about $50\%$. Fluorescence detection with a high-numerical-aperture objective enables single-atom readout with a fidelity of $99.986(6)\%$, while the achieved temperature of $\qty{12.92\pm0.05}{\micro\kelvin}$, survival probabilities of $97(2)\%$, and a vacuum lifetime exceeding $\qty{400}{\second}$ confirm the stability of the system. 
Together, these results establish a robust Sr tweezer platform~\citep{Gavryusev2024, Guariento2025} that combines efficient preparation, robust control, and high-fidelity detection, paving the way for scalable experiments on quantum simulation and quantum information processing with alkaline-earth atoms.

\section{Experimental Platform}
\label{sec:exp}

We realized a platform for trapping and manipulating single \isotope[88]{Sr} atoms in optical tweezers. The apparatus is based on a custom ultra-high vacuum (UHV) system that ensures long atomic lifetimes required for high-precision experiments. The science chamber is a fused silica glass cell providing excellent optical access, surrounded by two sets of coils. A pair of high-field coils supplies the magnetic field gradients for magneto-optical trapping as well as strong bias fields for clock-state manipulation, while six auxiliary compensation coils enable the three-dimensional control of the magnetic environment, allowing both active cancellation of stray fields and the application of tunable bias fields for state preparation.
A high-numerical aperture microscope objective simultaneously generates sub-micron waist optical tweezers and collects atomic fluorescence with single-atom resolution. This configuration combines precise control over individual atoms with high-fidelity state detection. Dedicated laser systems provide cooling, optical pumping, and trapping, completing a versatile platform ideally suited for scalable quantum simulation and quantum information processing with alkaline-earth atoms.

\subsection{Vacuum System}
\label{sec:exp:vac}

\begin{figure}[H]
    \centering
    \includegraphics[width=1\linewidth]{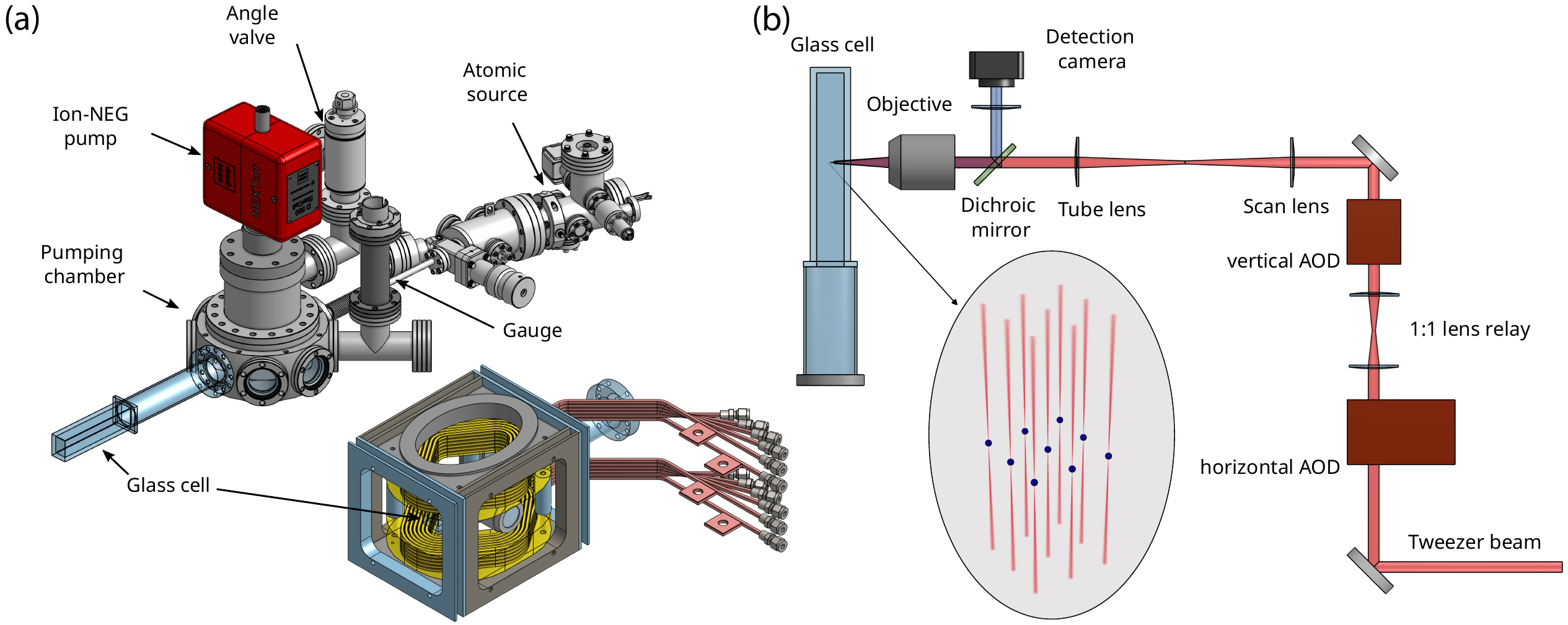}
    \caption{Overview of the experimental setup. (\textbf{a}) Vacuum system 
    composed of a commercial atomic source, connected through a differential pumping tube to a pumping chamber which holds an ion--NEG pump, a Bayard--Alpert gauge and an angle valve. The fused silica glass science cell, where trapping and detection occur, is surrounded by two sets of magnetic coils: a pair of high-field coils providing the MOT gradient and strong bias fields for clock-state manipulation, and six smaller compensation coils that enable full three-dimensional control and active cancellation of stray fields. 
    (\textbf{b}) Optical layout for tweezer generation and imaging. A high-power $\qty{813}{\nano\meter}$ beam passes through a pair of crossed AODs to produce an array of deflected beams. A relay telescope and a high-numerical-aperture objective (NA = 0.55) couple the angular deflection to the focal plane of the objective to form an array of spatially-displaced tightly-focused optical tweezers. The same objective collects atomic fluorescence, which is separated by a dichroic mirror and imaged onto a low-noise qCMOS detection camera for single-atom readout.}
    \label{fig:setup}
\end{figure}

The vacuum system is designed to maintain ultra-high vacuum (UHV) conditions, which are essential for trapping and manipulating single Sr atoms~\citep{Covey2019, Urech2022, Young2020a}. It consists of three main sections: a commercial atomic source, a pumping chamber, and a glass cell where the optical tweezers are formed and single-atom imaging takes place, as illustrated in Figure~\ref{fig:setup}a. A stable flux of \isotope[]{Sr} atoms is generated by an atomic source system (AOSense Inc., Fremont, USA), which includes a Zeeman slower as well as a 2D MOT stage that employ a permanent-magnet assembly to transversely cool and collimate the atomic beam. The slowed atoms are directed towards the science cell through a differential pumping tube, which preserves the vacuum gradient while allowing for a high flux of slowed atoms.
The 2D MOT acts also as an optically controlled deflection stage that allows for switching on and off the slowed-down atom flux, which is deflected by $20^\circ$ with respect to the Zeeman slower axis.
Following the differential pumping tube, the atoms pass through a dedicated pumping chamber that houses an Ion–NEG combination pump (SAES Getters S.p.A., Milano, Italy, NEXTorr D500 Starcell), a Bayard–Alpert UHV gauge (Agilent Technologies Inc., Santa Clara, USA, UHV-24P), and an angle valve that provides access to an external rough pumping station. During initial pump-down, a scroll pump and turbo pump are attached through this valve, while a residual gas analyzer (Stanford Research Systems, Sunnyvale, USA, RGA200) monitors partial pressures during the bakeout procedure. The pumping chamber was designed to minimize outgassing and ensure long-term pressure stability. Downstream, the atoms enter the rectangular science cell, whose elongated planar faces maximize optical access for the laser beams required for cooling, trapping, and imaging (anti-reflection coated on the wider sides). The entire vacuum assembly is mounted on a translation stage, which provides mechanical decoupling from the rest of the apparatus and facilitates alignment and maintenance.
The system was subject to an extended bakeout sequence to remove residual gases and routinely achieves base pressures below $\qty{7e-12}{\torr}$. Particular attention was devoted to material selection, sealing methods, and thermal cycling in order to suppress leaks and guarantee long-term integrity. Under these conditions, the vacuum system supports atom vacuum lifetimes compatible with long trapping and imaging sequences, as characterized in Section~\ref{sec:results:tweez}. This architecture thus provides the stable and collision-free environment required for high-fidelity single-atom trapping and detection.

\subsection{Magnetic Field Control}
\label{sec:exp:magn}

Two sets of coils are employed to control the magnetic environment of the experiment. The first consists of a pair of copper hollow-core water-cooled high-field coils (total resistance $\qty{60}{\milli\ohm}$ and inductance $\qty{600}{\micro\henry}$) capable of generating magnetic fields up to $\qty{1000}{\gauss}$ in a Helmholtz configuration. These coils provide the strong bias fields required for excitation of the clock transition in bosonic \isotope[88]{Sr}~\citep{Origlia2018, Baillard2007} and can also be operated in anti-Helmholtz configuration to generate the field gradients needed for the MOTs, reaching up to $\qty{100}{\gauss\per\centi\meter}$ at the peak current of $\qty{200}{\ampere}$ provided by the power supply (Delta Elektronika, Zierikzee, Netherlands, SM6000 30-200). In addition, three pairs of low-field compensation coils, driven by smaller power supplies (Delta Elektronika, ES150 015-10), are used to provide three-dimensional compensation of stray fields and fine adjustment of the trapping region in the red MOT. 

To achieve rapid magnetic field switching, we implemented a fast discharge \textit{snubber} circuit using a solid state electronic design~\citep{Guariento2025} with time constant of $\qty{240\pm10}{\micro\second}$. An Insulated-Gate Bipolar Transistor (IGBT) (Semikron, Nuremberg, Germany, SKM400GAL12T4) controls the current flow in the high-field coils and is actuated by a gate driver (Semikron, SKHI 10/17R) triggered via a TTL signal. When the IGBT is turned off, the inductive energy loaded into the coils has to be dissipated and this causes a voltage surge within the circuit that can reach hundreds of volts on a sub-millisecond timescale. Voltage surge protection is provided by placing a varistor (LittleFuse, Rosemont, USA, V25S250P, conduction voltage $\geq\qty{390}{\volt}$) in parallel with the IGBT, and, in parallel with the coils, a branch with a diode and a $\qty{10}{\ohm}$ resistor in series, to ensure safe and fast dissipation when the varistor is below threshold. An additional protective varistor (conduction voltage $\geq\qty{75}{\volt}$) is located in parallel with the power supply. This circuit allows a fast current decay in $\qty{240\pm10}{\micro\second}$ that minimizes residual magnetic fields, essential for stage transitions (e.g., blue-to-red MOT as depicted in figure~\ref{fig:sequence}) and for precise imaging free from Zeeman shifts.
As a future improvement, a custom H-bridge circuit has been designed and will soon be implemented, enabling fast switching between Helmholtz and anti-Helmholtz operation~\citep{Madjarov2020}.

\subsection{Laser Systems}
\label{sec:exp:laser}

The experimental setup relies on four primary laser systems that provide cooling, trapping, and coherent manipulation of individual~\isotope[88]{Sr} atoms. The electronic level structure of this isotope is presented in Figure~\ref{fig:repumpers}(a), displaying the wavelength and the decay rate of the transitions of interest.
A blue laser at $\qty{461}{\nano\meter}$ (Moglabs, Carlton, Australia, ILA) addresses the broad $\ket{^1\mathrm{S}_0} \rightarrow \ket{^1\mathrm{P}_1}$ transition and serves multiple purposes in the experimental sequence: it provides Zeeman slowing of the atomic beam, drives the 2D MOTs, operates the first stage of the 3D MOT, and is used for fluorescence imaging. The two associated \textit{repumping} transitions ($\ket{^3\mathrm{P}_0} \rightarrow \ket{^3\mathrm{S}_1}$ at $\qty{679}{\nano\meter}$ and $\ket{^3\mathrm{P}_2} \rightarrow \ket{^3\mathrm{S}_1}$ at $\qty{707}{\nano\meter}$) are addressed by two tunable diode lasers (Toptica Photonics AG, Munich, Germany, DL Pro).
A second stage of 3D MOT is performed with a red laser at $\qty{689}{\nano\meter}$ (MSquared, Glasgow, UK, SolsTiS-PSX-PIK), resonant with the narrow intercombination line $\ket{^1\mathrm{S}_0} \rightarrow \ket{^3\mathrm{P}_1}$, that is also used for in-tweezer cooling and manipulation. Optical tweezers are formed by an $\qty{8}{\watt}$ near-infrared laser at $\qty{813}{\nano\meter}$ (Sirah, Grevenbroich, Germany, Matisse 2 TS), delivered to the experiment table through a photonic crystal fiber (NKT Photonics, Regensdorf, Switzerland, LMA-PM-15) and operating at the magic wavelength for the $\ket{^1\mathrm{S}_0} \rightarrow \ket{^3\mathrm{P}_0}$ clock transition, thereby suppressing differential light shifts.
In addition, ultraviolet light in the $316-\qty{319}{\nano\meter}$ range is generated by frequency quadrupling a tunable infrared source (Vexlum, Tampere, Finland, VALO SF $1264-\qty{1278}{\nano\meter}$) in a two-stage fourth harmonic generation (FHG) cavity (LEOS Solutions, Rovereto, Italy), producing more than $\qty{110}{\milli\watt}$ of UV power. The wide tunability of this system will enable two complementary excitation schemes to $\ket{n\mathrm{S}}$ Rydberg states, either via the $\ket{^3\mathrm{P}_0}$ clock state in a two-step process~\citep{Madjarov2020} or through a two-photon transition from the $\ket{^3\mathrm{P}_1}$ state~\citep{Qiao2021}.

Frequency stabilization is adapted to the specific requirements of each laser. The $\qty{461}{\nano\meter}$ system is locked to the spectroscopic reference provided by the $\ket{^1\mathrm{S}_0} \rightarrow \ket{^1\mathrm{P}_1}$ transition, while the tweezer laser at $\qty{813}{\nano\meter}$ and the two repumper lasers are stabilized using the feedback of a high-resolution wavemeter (HighFinesse, Tubinga, Germany, WS8-10, $\qty{10}{\mega\hertz}$ absolute accuracy). For the narrow-linewidth transitions, both the $\qty{689}{\nano\meter}$ and the UV lasers (through the intermediate SHG emission at $632-\qty{638}{\nano\meter}$) are stabilized to an ultra-low expansion cavity (Stable Laser Systems) with high finesse ($\approx \qty{2.65e5}{}$) using Pound–Drever–Hall locking, yielding $\unit{\kilo\hertz}$ linewidths and frequency stability compatible with precision control of alkaline-earth atoms.

\subsection{Two-Stage Magneto-Optical Trap}
\label{sec:exp:mot}

To reach the temperatures and densities required for efficient atom loading into the optical tweezers we employ a two-stage MOT, which provides sufficiently low temperatures and high atomic densities~\citep{Sorrentino2006, Norcia2018a, Snigirev2019, Wen2024}. The atomic beam is first slowed and transversely cooled using a Zeeman slower and a two-dimensional MOT stage inside the AOSense assembly, providing a continuous flux of atoms with an average speed of $\qty{40}{\meter\per\second}$. These atoms are then captured in a two-stage, three-dimensional MOT.

The first \textit{blue} MOT operates on the broad $\ket{^1\mathrm{S}_0} \rightarrow \ket{^1\mathrm{P}_1}$ transition at $\qty{461}{\nano\meter}$ ($\Gamma/2\pi = \qty{32}{\mega\hertz}$). During this stage, two repumping lasers at $\qty{679}{\nano\meter}$ and $\qty{707}{\nano\meter}$, addressing the $\ket{^3\mathrm{P}_0} \rightarrow \ket{^3\mathrm{S}_1}$ and $\ket{^3\mathrm{P}_2} \rightarrow \ket{^3\mathrm{S}_1}$ transitions, respectively, prevent population trapping in dark states~\citep{Shaokai2009, Hu2019, Patel2024}, allowing efficient capture of atoms and cooling to a few $\unit{\milli\kelvin}$, while increasing the atom number by approximately an order of magnitude. 

However, density-dependent losses and radiation trapping in this stage limit the achievable atom number~\citep{Hoeschele2023}. To mitigate these effects, we apply a weak beam resonant with the fully cycling narrow intercombination line $\ket{^1\mathrm{S}_0} \rightarrow \ket{^3\mathrm{P}_1}$ at $\qty{689}{\nano\meter}$ ($\Gamma/2\pi = \qty{7.6}{\kilo\hertz}$) that transfers atoms into a long-lived state, effectively reducing reabsorption and inelastic collisions in the blue MOT volume, such that it acts as a loss \textit{shield}. As demonstrated in recent studies~\citep{Hoeschele2023}, this mechanism enhances the steady-state atom number up to a factor of $2$ by providing an auxiliary reservoir that continuously replenishes the blue MOT, while suppressing light-assisted losses. As the $\qty{689}{\nano\meter}$ laser system is already employed in the second MOT stage, the inclusion of the \textit{shielding} beam does not add experimental complexity.

Considering that temperatures and densities achieved in the first MOT stage are insufficient for direct tweezer loading, a second \textit{red} cooling stage is implemented using the narrow intercombination line $\ket{^1\mathrm{S}_0} \rightarrow \ket{^3\mathrm{P}_1}$. Since the linewidths of the blue and red transitions differ by more than three orders of magnitude, the radiation pressure force from the narrow-line red MOT is too weak to efficiently capture and cool atoms moving at velocities of order $\sim \unit{\meter\per\second}$, as is typical for atoms in the blue MOT. To overcome this limitation, we perform an intermediate \textit{multi-frequency broadband} (BB) red MOT stage in which the $\qty{689}{\nano\meter}$ light is spectrally broadened by modulating, with an arbitrary waveform generator (AWG) (Siglent Technologies, Shenzhen, China, SDG6022X), the radio-frequency (RF) signal that drives an acousto-optical modulator (AOM)~\citep{Poli2014, Urech2022}. This loading phase increases the effective scattering rate over a wider velocity range, which enables capturing most of the atoms from the blue MOT and bridge the large gap between the Doppler cooling regimes of the blue and red transitions~\citep{Phalen2005}. 
Once the atoms are transferred, the broadband modulation is turned off and the \textit{single-frequency} (SF) red MOT provides final cooling to the few-$\unit{\micro\kelvin}$ regime, while concurrently raising the atomic density.

Both the blue and red MOTs are formed by three retro-reflected beams, a configuration chosen to simplify the optical setup and reduce laser power requirements compared to six independent beams.
The two horizontal beams have a high incidence angle of $\approx \qty{65}{\degree}$ on the glass cell surface, which leads to a $16\%$ power loss for the reflected beams, that is compensated by refocusing them with a long-focal-distance lens (Thorlabs, Newton, USA, LA1259-AB). 
The MOTs are characterized using absorption imaging. A resonant $\qty{461}{\nano\meter}$ probe beam is imaged onto a CMOS camera (Basler, Ace acA4024-29um), and a standard sequence of three images (atoms, probe beam, and background) is acquired. The background is subtracted from both the atoms and probe images, and then the ratio of the latter two is used to retrieve a 2D map of the optical column density (OD) profile that, through a calibration, allows the precise determination of the atom number and of the cloud density.

\subsection{Optical Tweezer Array and Single-Atom Detection}
\label{sec:exp:tweez}

The optical tweezer array is generated by modulating a high-power $\qty{813}{\nano\meter}$ beam with a pair of orthogonally oriented acousto-optic deflectors (AODs) (AA Opto-Electronic, Orsay, France, DTSX-400-810)~\citep{Labuhn2014, Nogrette2014, Barredo2016, Ricci2022, Ricci2024}, coupled by a $1:1$ lens relay in a $4f$ arrangement, as illustrated in Figure~\ref{fig:setup}b. The RF signals that drive the AODs can be actively programmed by an AWG (Spectrum Instrumentation, Grosshansdorf, Germany, M4i.6631-x8). This configuration provides dynamic control over the beam positions and amplitudes, enabling the creation of programmable tweezer patterns with adjustable geometry and site spacing. The modulated light is then tightly focused onto the atomic ensemble using a custom high-numerical-aperture objective (Special Optics, NA = 0.55), mounted on a five-axis positioning stage, producing optical traps with sub-$\mu$m waist and suitable depth for single-atom confinement.

The atoms loaded into the tweezers undergo further cooling via narrow-line Sisyphus cooling~\citep{Dalibard1989, Urech2022, Hoelzl2023} on the intercombination transition. To ensure single-atom occupancy, light-assisted collisions (LAC)~\citep{Sompet2013, Labuhn2016a, Fung2016a, Cooper2018, Brown2019, Jenkins2022} are induced, which efficiently remove pairs of atoms from the traps, while preserving single atoms (see Section~\ref{sec:results:tweez}). This process yields an average single-site filling fraction of approximately $50\%$, consistent with the stochastic nature of collisional-blockade loading.

High-fidelity detection of the trapped atoms is achieved using the same high-NA objective employed for tweezer generation~\citep{Cooper2018, Endres2016, Tao2024}. Atomic fluorescence, collected during resonant orthogonal illumination on the $\qty{461}{\nano\meter}$ transition, is imaged onto a low-noise qCMOS camera (Hamamatsu Photonics, Hamamatsu, Japan, ORCA-Quest). The optical system provides high resolution at the single-site level, while the camera's sensitivity and low background noise enable single-atom detection with a high fidelity exceeding $99.9\%$.
This combination of programmable optical potentials, efficient cooling, controlled collisional loading, and high-fidelity imaging are the key features required to make our experimental setup a robust and scalable platform for quantum simulation and quantum information processing with alkaline-earth atoms.

\subsection{Experiment Control System}
\label{sec:exp:control}

To guarantee an efficient, flexible and reliable experiment preparation and execution, the entire experimental cycle is controlled through the \textit{Labscript Suite}~\citep{Starkey2013}, an open-source Python-based framework specifically designed for atomic-physics experiments. It provides intuitive graphical user interfaces (GUI) and on-line data analysis, with closed-loop feedback and optimization capability. 
Several devices are natively supported by the Labscript developers (e.g., PulseBlaster boards, IMAQ cameras), while additional hardware support has been contributed by users over time. In our experiment, the currently implemented devices include a custom FPGA timing system~\citep{Trenkwalder2021}, based on the Cora Z7 Zynq-7000 System on a Chip, providing through extension boards many TTL and $\qty{\pm10}{\volt}$ analog outputs; Moglabs QRF four channel radio-frequency drivers for AOM/AOD; Basler Ace cameras; Andor iXon Ultra 888 and Hamamatsu ORCA-Quest cameras; and Spectrum Instrumentation M4i AWG boards.

\begin{figure}[H]
    \centering
    \includegraphics[width=1\columnwidth]{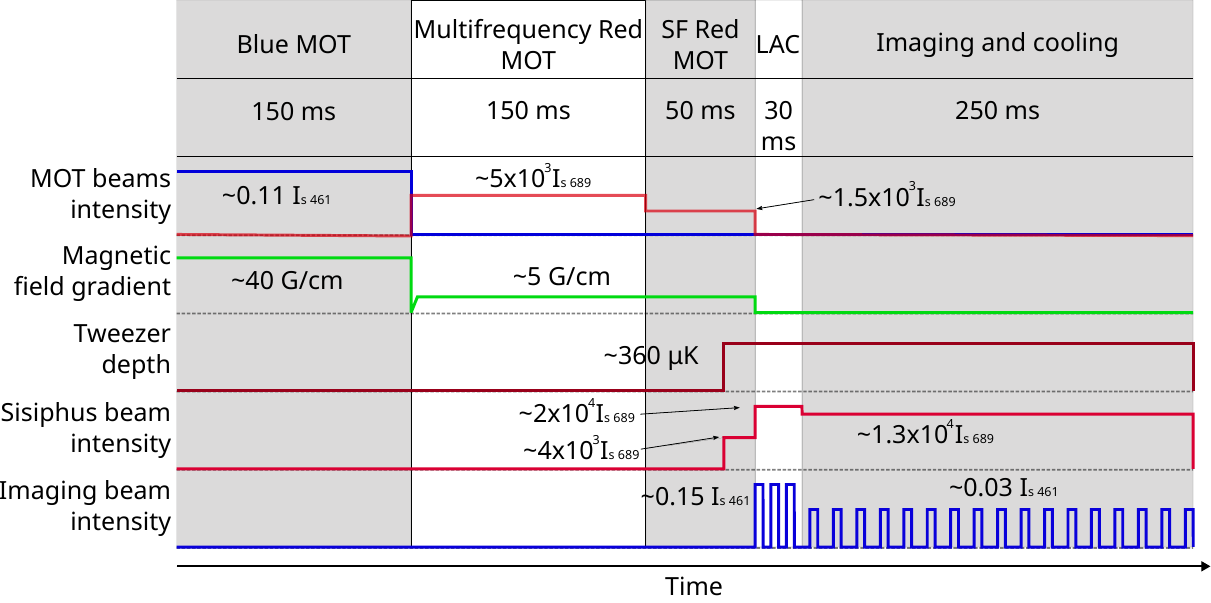}
    \caption{Temporal sequence of the atomic cooling, trapping in tweezers and imaging process. The diagram shows the different experimental stages (Blue MOT, Multifrequency Red MOT, Single Frequency Red MOT, LAC, Imaging and cooling) with their durations, along with the time evolution of the optical beam intensities, magnetic field gradient, and tweezer potential depth.}
    \label{fig:sequence}
\end{figure}

The sequences that control each specific experimental run are written as a single Python script. To streamline sequence development, we created a dedicated library that provides functions for common experimental actions, such as TTL pulses, voltage ramps, or even for entire routines, such as MOT loading or imaging procedures, making the script sequence easy to be read. This library can be simply imported into the script and the execution of each sequence block can be configured via case flags that are set through boolean values in the global parameters section of the \textit{runmanager} GUI. Therefore, from the same interface, users can easily adjust parameters, launch scans, and perform calibrations. Figure~\ref{fig:sequence} displays a typical experimental run, consisting of five blocks corresponding to the three atomic cooling stages, optical trapping in tweezers and  imaging. Their respective durations, optical beam intensities, magnetic field gradient, and tweezer potential depth are also shown.

\section{Characterization and Results}
\label{sec:results}

\subsection{Absorption Imaging Analysis Procedure}
\label{sec:results:analysis}

The MOTs are characterized using standard absorption imaging and the atom number $N_\text{atoms}$ is extracted by fitting the absorption profile with a two-dimensional Gaussian (with $\sigma_{x}^2,\,\sigma_{y}^2$ variances), which after integration yields
\begin{linenomath}
\begin{equation}
N_\text{atoms}=2\pi \ \sigma_{x} \sigma_{y}\ \frac{n_0}{\sigma} ,
\end{equation}
\end{linenomath}
where $n_0$ is the optical density peak and $\sigma$ is the atomic resonant absorption cross section.

All absorption images are acquired in time-of-flight (TOF) measurements, where  the MOT lasers and magnetic fields are switched off and the atoms then expand freely under gravity.  The expansion dynamics of the cloud is determined by its temperature. For a thermal gas far from quantum degeneracy, the velocity distribution follows the Maxwell–Boltzmann statistics~\citep{Metcalf1999}, giving
\begin{linenomath}
\begin{equation}
\tilde{n}(x,y,t)= n_o\exp{\left[ -\frac{(x-x_0)^2}{2\sigma_x^2(t)}-\frac{(y-y_0)^2}{2\sigma_y^2(t)} \right]}, \quad \sigma_i^2(t)= \sigma_{0i}^2+\sigma_v^2t^2,
\end{equation}
\end{linenomath}
with $\sigma_v^2= k_B T/M$, where $M$ is the atomic mass. The cloud widths  $\sigma_i(t)$ increase with the expansion time $t$, and by fitting $\sigma_x(t)$ and $\sigma_y(t)$ we obtain independent estimates of the expansion velocity. Within the experimental uncertainty, we find $\sigma_{v_x}^2=\sigma_{v_y}^2$, and use their average $\sigma_{avg}(t)$ to extract the temperature.

\subsection{Blue Shielded MOT}
\label{sec:results:motblue}

\begin{figure}[H]
    \centering
    \includegraphics[width=1\linewidth]{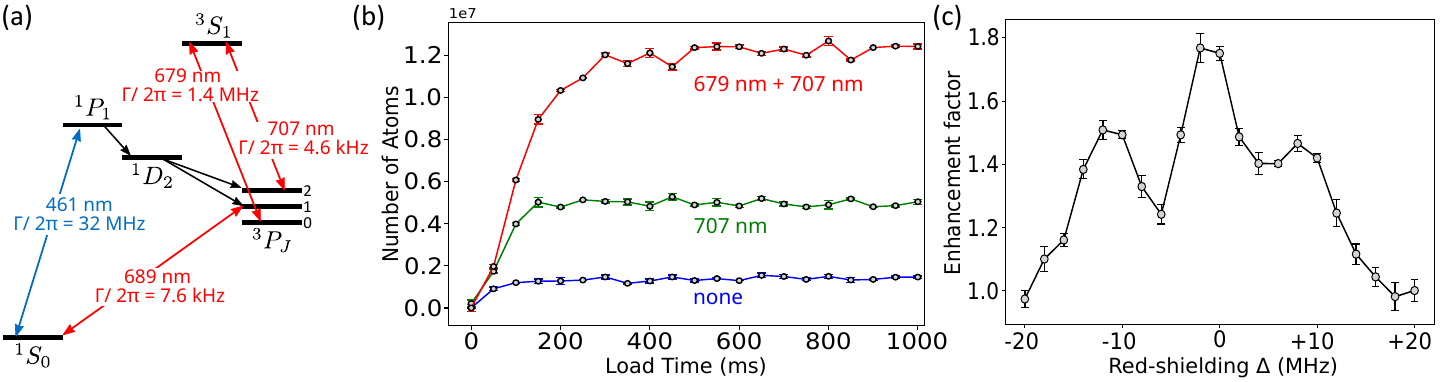}
    \caption{Blue shielded MOT characterization. (\textbf{a}) Simplified electronic level structure of~\isotope[88]{Sr} displaying the transitions of interest. (\textbf{b}) Loading curves of the shielded blue MOT for different repumping configurations. The presence of the $\qty{707}{\nano\meter}$ repumper (green) increases the number of trapped atoms by preventing optical pumping into the dark metastable $\ket{^3\mathrm{P}_2}$ state. When the $\qty{679}{\nano\meter}$ repumper is also simultaneously applied (red), the loss channel to the dark metastable  $\ket{^3\mathrm{P}_0}$ state is also inhibited, which increases the atom number by more than one order of magnitude compared to the case without any repumper (blue). (\textbf{c}) Relative enhancement of the atom number by the red-shielding beam as a function of the frequency detuning from resonance, assuming the far from resonance values as baseline. Error bars in (\textbf{b},\textbf{c}) represent standard deviations.}
    \label{fig:repumpers}
\end{figure}

Under our typical operating conditions, we obtain a blue MOT collecting around $\qty{1.25e7}{}$ atoms at a peak density of $\qty{4e9}{\per\cubic\centi\meter}$ over $\qty{150}{\milli\second}$, with a temperature of $\qty{6.8}{\milli\kelvin}$. Its performance is optimized by scanning key parameters, including the quadrupole magnetic field gradient, the power and detuning of the 3D MOT beams, and the power and detuning of the 2D MOT and Zeeman slower beams. 
Figure~\ref{fig:repumpers}a presents a simplified schematic of the electronic level structure of \isotope[88]{Sr}, showing the key optical transitions employed during this stage of the experiment.

In Figure~\ref{fig:repumpers}b, we report the characterization of the dynamics of the blue \textit{shielded} MOT cloud loading with/without the repumping lasers. The presence of the $\qty{707}{\nano\meter}$ repumper (green) triples the number of trapped atoms by preventing optical pumping into the dark metastable $\ket{^3\mathrm{P}_2}$ state. When the $\qty{679}{\nano\meter}$ repumper is also simultaneously applied (red), the loss channel to the dark metastable  $\ket{^3\mathrm{P}_0}$ state is also inhibited, which increases the atom number by more than one order of magnitude compared to the case without repumpers (blue). Synergistically, the use of the \textit{shielding} beam yields a $\approx 80\%$ improvement in atom number with respect to the unshielded configuration, as shown in Figure~\ref{fig:repumpers}c. The spectrum shape is due to the Zeeman splitting of the different $m_J$ Zeeman substates of $\ket{^3\mathrm{P}_1}$ integrated over the MOT extension~\citep{Hoeschele2023}.

\begin{figure}[H]
    \centering
    \includegraphics[width=0.9\linewidth]{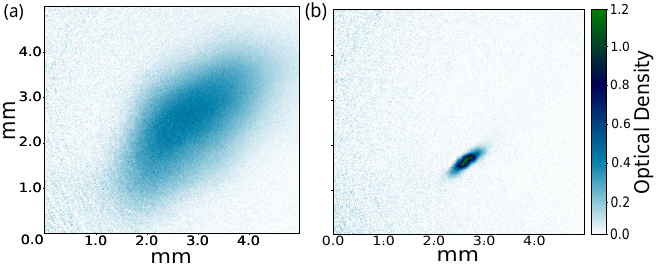}
    \caption{Absorption images of laser-cooled \isotope[88]{Sr} clouds: (\textbf{a}) a blue shielded MOT with $\sim 1.25 \times 10^7$ atoms at $\qty{6.8}{\milli\kelvin}$ and  
    (\textbf{b}) a single-frequency red MOT with more than $5\times10^{5}$ atoms at $\qty{5.4}{\micro\kelvin}$. The false-color optical density scale is the same in both absorption images to allow direct visual comparison. }
    \label{fig:motimg}
\end{figure}

To account for correlations between parameters, such as laser power and detuning $\Delta$, two-dimensional optimization scans (not reported as figures) are performed. Complementarily, we report the saturation parameter $s=I/I_\text{sat}$, where $I$ is the peak beam intensity and $I_\text{sat}$ is the saturation intensity for the addressed transition. Optimal conditions are found for 2D MOT beams with powers of $\qty{9.4}{\milli\watt}$ ($s=0.11$) and $\qty{4.7}{\milli\watt}$ ($s=0.055$) at the entrance of the atomic source chamber, detuned by $\Delta_\text{2D,461}/2\pi = -0.7\,\Gamma_{^1\mathrm{P}_1}/2\pi = \qty{-22}{\mega\hertz}$, with saturation intensity $I_\text{sat,461}= \qty{42}{\milli\watt\per\square\centi\meter}$.
The Zeeman slower optimal operation power is $\qty{40}{\milli\watt}$ ($s=0.93$) with a detuning of $\Delta_\text{Zeeman,461}/2\pi = -19\,\Gamma_{^1\mathrm{P}_1}/2\pi= \qty{-580}{\mega\hertz}$. 
The 3D MOT beams operate optimally at $\qty{4.3}{\milli\watt}$ per beam ($s=0.1$) with a detuning of $\Delta_\text{3D,461}/2\pi = -2\,\Gamma_{^1\mathrm{P}_1}/2\pi = \qty{-60}{\mega\hertz}$. The quadrupole gradient is set to $\qty{50}{\gauss\per\centi\meter}$, which provides the best compromise between atom number and density. A typical absorption image of the blue shielded MOT is shown in Figure~\ref{fig:motimg}a.

\subsection{Multi-Frequency Red MOT}
\label{sec:results:motredmf}

For the intermediate cooling stage, we employ a BB red MOT for $\qty{150}{\milli\second}$. To generate a multi-frequency laser beam, we modulate the RF signal driving the AOM that controls the red 3D MOT beams. An AWG (Siglent Technologies, Shenzhen, China, SDG6022X) produces a sinusoidal modulation of the RF carrier with frequency $\qty{30}{\kilo\hertz}$ and modulation depth $\qty{\pm2.4}{\mega\hertz}$. The Fourier spectrum of this signal consists of a comb of frequency components with $\qty{30}{\kilo\hertz}$ spacing, extending over $\qty{4.8}{\mega\hertz}$. Owing to the double-pass configuration of the AOM, the effective frequency comb in the laser light covers $\qty{9.6}{\mega\hertz}$. Among the available modulation schemes, the sinusoidal modulation provides the steepest roll-off of sidebands outside the modulation span, ensuring that the comb terminates near the atomic resonance without introducing significant blue-detuned components that could otherwise heat the atoms. Ideally, the comb should be fully red-detuned, extending from few $\Gamma_{^3\mathrm{P}_1}$ below the $^1\mathrm{S}_0$ $\to$ $^3\mathrm{P}_1$ resonance toward lower frequencies. In practice, with our AWG we can adjust the central frequency and total span, which set the effective upper and lower limits of the comb. Immediately after the transfer from the blue MOT, all frequency components contribute to the cooling process. As the atoms decelerate, the Doppler shift decreases and only the less-detuned frequencies remain effective. In principle, the frequency comb could be dynamically narrowed to a single frequency to further cool the atoms. Nevertheless, as we proved, efficient transfer into a single-frequency red MOT is still possible because the saturated linewidth covers the gap between the broadband comb and a narrowband single frequency. Each 3D MOT beam in the BB stage has $\qty{4}{\milli\watt}$ total power distributed over 320 frequency components. For a beam waist of $\qty{8}{\milli\meter}$ and a saturation intensity of $I_\text{sat,689} = \qty{3}{\micro\watt\per\centi\meter\squared}$, we estimate that each component has a saturation parameter of $s \simeq 4.2$, corresponding to a power-broadened linewidth of $\Gamma_s / 2\pi = \qty{17}{\kilo\hertz}$. 

During the transfer from the blue to the red MOT, the magnetic field gradient must be rapidly reduced from $\qty{50}{\gauss\per\centi\meter}$ to a few $\unit{\gauss\per\centi\meter}$. This is implemented by switching off the quadrupole coils at the end of the blue 3D MOT stage, together with the blue 3D MOT beams, and then reactivating them at the desired current value. The fast current switch-off circuit enables a decay time of $\qty{240\pm10}{\micro\second}$, leaving the atoms in free fall for just a few hundred microseconds before the current ramps up again. As the gradient increases from $0$ to the set value, the trapping region contracts and the cloud is gradually compressed. After testing different final values, we selected a gradient of $\qty{5}{\gauss\per\centi\meter}$, which yields the largest trapped atom number and highest density. A TOF measurement during the broadband stage gives a temperature of $T_\text{BB} = \qty{21}{\micro\kelvin}$ with approximately $\qty{5e5}{}$ atoms trapped with a peak density of $\qty{3.5e8}{\per\centi\meter\cubed}$.

\subsection{Single-Frequency Red MOT}
\label{sec:results:motredsf}

The final cooling stage is a SF red MOT that lasts for $\qty{50}{\milli\second}$. Efficient transfer from the BB MOT is achieved by starting the SF sequence at a high laser intensity saturation parameter ($s \simeq 200$), which corresponds to a power-broadened linewidth of $\Gamma_s/2\pi = \qty{105}{\kilo\hertz}$. This broadening bridges effectively the frequency gap between the broadband comb and a narrow single frequency. With this approach, more than $80\%$ of the atoms are transferred from the BB to the SF MOT. Then the power is linearly ramped down by $20\%$. A TOF expansion yields a final temperature of $\qty{5.4}{\micro\kelvin}$, compared to the Doppler limit of $\qty{2.4}{\micro\kelvin}$ for the broadened linewidth. Importantly, only in this stage we achieve atomic densities above the threshold required for efficient tweezer loading ($n > \qty{3e10}{\per\centi\meter\cubed}$, estimated assuming an average of one atom in a tweezer trapping volume). Under optimized conditions, we obtain more than $\qty{5e5}{}$ atoms at a density of $n \simeq \qty{4e10}{\per\centi\meter\cubed}$. A typical SF MOT is shown in Figure~\ref{fig:motimg}b and has waists $\sigma_x = \qty{50}{\micro\meter}$ and $\sigma_y = \qty{250}{\micro\meter}$. To maintain optimal conditions and compensate for the slow linear drift of the ULE cavity used for frequency stabilization ($\sim \qty{12}{\kilo\hertz/day}$), we perform daily optimization scans of the SF MOT laser frequency and power, typically achieving the best performance at a detuning $\Delta_\text{3D,689}/2\pi = \qty{-100}{\kilo\hertz}$ and $s \simeq 200$.

\subsection{Single Atoms in Optical Tweezers Array}
\label{sec:results:tweez}

The AOD-generated optical tweezer array is turned on shortly before the end of the red SF MOT and can span a two-dimensional field of view with a lateral size of $\sim \qty{400}{\micro\meter}$, which is comparable to the atom cloud size. Using the compensation coils, we achieve precise spatial overlap between the MOT and the tweezers array, with a sensitivity of $\qty{2}{\micro\meter}/\qty{}{\milli\gauss\per\centi\meter}$. Each optical tweezer has a waist of $\sim\qty{1.5}{\micro\meter}$ and a trap depth of $\sim\qty{360}{\micro\kelvin}$, well suited for single-atom trapping. 

To achieve single-atom occupancy, we apply a LAC sequence for $\qty{30}{\milli\second}$ by shining a $\Delta_\text{LAC,689}/2\pi=\qty{-2}{\mega\hertz}$ red-detuned $\qty{689}{\nano\meter}$ beam ($s=\qty{2e4}{}$) and we speed up the loss process by adding pulsed resonant $\qty{461}{\nano\meter}$ light with a $\qty{1}{\milli\second}$ duration and $\qty{333}{\hertz}$ repetition rate ($s=0.15$), as in~\citep{Norcia2018}. As a result, we obtain a single-atom loading probability of $\sim50\%$ across the array.

\begin{figure}[H]
  \centering
    \includegraphics[width=1\columnwidth]{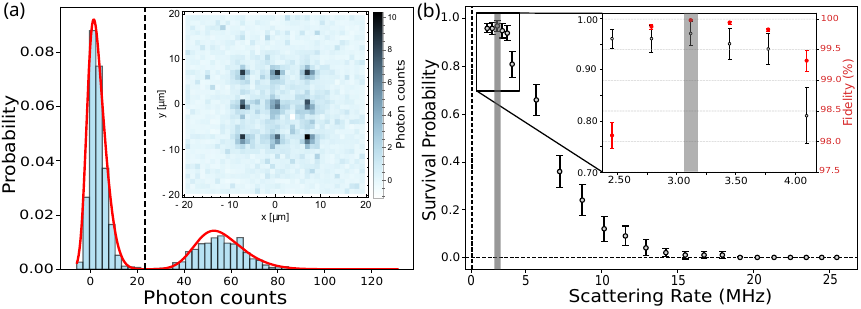}
    \caption{Characterization of single atoms in tweezers through fluorescence detection. (\textbf{a}) The histogram of photon counts collected in ROIs centered at the atom positions (averaged over the array) features a bimodal distribution corresponding to empty and occupied traps, that is fitted with the sum of two skewed gaussians (red line); the dashed line at $\sim 23$ photon counts indicates the threshold value that maximizes the fidelity of true positive single-atom detection.
    Insert: typical fluorescence image of a $3\times3$ array of single atoms, obtained by averaging 100 individual acquisitions. 
   (\textbf{b}) Trapped atom survival probability and detection fidelity $\mathcal{F}$ (in red in the inset that focuses on the first points) as a function of the probe beam scattering rate. All data is taken with a fixed imaging duration of $\qty{150}{\milli\second}$, chosen to ensure that for every scattering rate the detected signal was sufficiently strong to allow accurate evaluation of both quantities. The shaded region indicates the optimal scattering-rate value. Error bars represent standard deviations calculated with the bootstrapping method~\citep{Efron1993}.}
   \label{fig:fidelity}
\end{figure}

Next, to detect the tweezer occupation, fluorescence imaging is performed for $\qty{250}{\milli\second}$ by pulsing a resonant $\qty{461}{\nano\meter}$ beam ($\qty{1}{\milli\second}$ pulses at $\qty{333}{\hertz}$ repetition rate, $s=0.03$). During imaging, a $\Delta_\text{Sis,689}/2\pi=\qty{-2.1}{\mega\hertz}$ red-detuned $\qty{689}{\nano\meter}$ beam ($s=\qty{1.3e4}{}$) is used to implement Sisyphus cooling on the $\ket{^1\mathrm{S}_0} \rightarrow \ket{^3\mathrm{P}_1}$ intercombination line. This cooling stage mitigates photon-scattering-induced heating and enables repeated non-destructive imaging. An image averaged over $100$ realizations of a $3\times3$ array of single atoms is shown in the inset of Figure~\ref{fig:fidelity}(a).
The resulting bimodal photon-count distribution collected at the expected atom positions in a region of interest of $3\times3$ pixels (Fig.~\ref{fig:fidelity}(a)) allows us to discriminate between empty and occupied tweezers.
The histogram of the photon distribution is fitted with the sum of two skewed gaussians representing, respectively, the positive detection of an empty site and of a singly occupied one. We define a threshold that determines whether a detected number of counts is attributed to the two situations. 
The tail of the empty trap distribution that lies beyond the threshold represents the false positive error probability $\varepsilon_1$, while the tail of the occupied trap distribution that lies below the threshold represents the false negative error probability $\varepsilon_2$. The fidelity is calculated as $\mathcal{F}=1-(\varepsilon_1+\varepsilon_2)$ and the threshold value is optimized to maximize $\mathcal{F}$, finding a peak single-atom detection fidelity of $\mathcal{F}=99.986(6)\%$ (Figure~\ref{fig:fidelity}b, red data in the inset that focuses on the first points).

Furthermore, the implementation of Sisyphus cooling ensures a survival probability of $97(2)\%$ during imaging, confirming the robustness of the protocol (Figure~\ref{fig:fidelity}b, black data). The survival probability is obtained from two consecutive images as the ratio between the number of populated tweezers in the second image over those detected in the first, providing a direct measure of atom loss during imaging.

\begin{figure}[H]
    \centering
    \includegraphics[width=1\columnwidth]{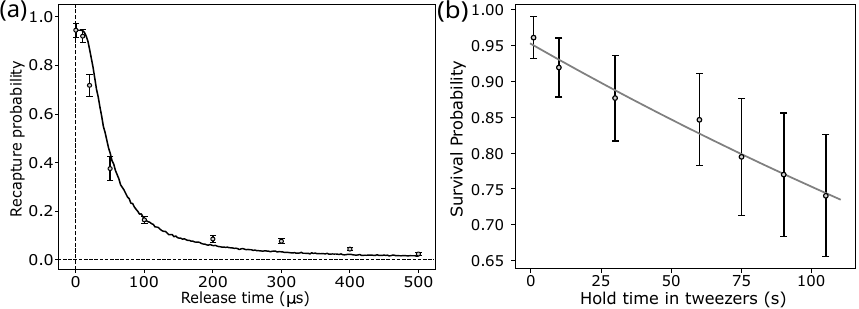}
    \caption{(\textbf{a}) Recapture probability as a function of release time during a release-and-recapture measurement, which provides an atom temperature of $\qty{12.92\pm0.05}{\micro\kelvin}$ through the comparison with Monte-Carlo simulations (solid line). 
    (\textbf{b}) Estimation of the vacuum-limited lifetime by fitting the atom survival probability as a function of the tweezer hold time with an exponential function (solid line). The fitted lifetime is $\qty{425\pm25}{\second}$. Error bars in (\textbf{a},\textbf{b}) represent standard deviations calculated with the bootstrapping method~\citep{Efron1993}.}
    \label{fig:lifetime}
\end{figure}

We characterize the temperature of the trapped atoms using a release-and-recapture measurement (Figure~\ref{fig:lifetime}a)~\citep{Tuchendler2008, MuziFalconi2025}. In this technique, the trapping potential is briefly switched off for a variable release time, after which the tweezers are restored and the probability of recapturing the atom is measured by acquiring a second image. As expected, the recapture probability decreases with longer release times, since atoms with higher kinetic energy are more likely to escape. The experimental data is compared to a Monte-Carlo simulation of atomic dynamics in the tweezer potential, from which we extract the atomic temperature, and we find the best agreement at a temperature of $\qty{12.92\pm0.05}{\micro\kelvin}$.

Finally, to assess the background vacuum conditions, we measure the lifetime of atoms trapped in the tweezers (Figure~\ref{fig:lifetime}b) by holding them for a variable time, up to several minutes, after which we acquire a second fluorescence image from which we extract the survival probability. The population decay is fitted to an exponential law, from which we extract a vacuum-limited lifetime of $\qty{425\pm25}{\second}$. This long lifetime indicates very good ultra-high vacuum conditions in the experimental chamber and demonstrates that background gas collisions are not a limiting factor for the experiments reported in this work.

\section{Conclusions}
\label{sec:concl}

We have developed and characterized a neutral-atom quantum simulation platform based on individually trapped \isotope[88]{Sr} atoms in optical tweezers. A key component of the system is a two-stage MOT operating on the $\ket{^1\mathrm{S}_0} \rightarrow \ket{^1\mathrm{P}_1}$ transition at $\qty{461}{\nano\meter}$ and on the narrow $\ket{^1\mathrm{S}_0} \rightarrow \ket{^3\mathrm{P}_1}$ intercombination line at $\qty{689}{\nano\meter}$. Using a blue shielded MOT scheme allows us to enhance the atom number by up to $80\%$. Sequential broadband and single-frequency cooling stages yield atomic samples at temperatures of $\qty{5.4}{\micro\kelvin}$ and densities $\qty{4e10}{\per\cubic\centi\meter}$, providing ideal conditions for tweezer loading.

The achieved single-atom detection fidelity of $\mathcal{F}=99.986(6)\%$ and survival probability of $97(2)\%$ are comparable to those reported in similar Yb and Sr tweezer experiments~\citep{Norcia2018, Cooper2018, Saskin2019, Wilson2022}, while the measured temperature of $\sim \qty{13}{\micro\kelvin}$ in the optical tweezers confirms the efficiency of our in-tweezer cooling scheme. 
Furthermore, the observed vacuum-limited lifetime exceeding seven minutes is on par with, or better than, similar optical tweezer systems~\citep{Madjarov2020, Jenkins2022}, highlighting the quality of the ultra-high-vacuum environment achieved in our setup. 
These performance indicators demonstrate that our apparatus provides the necessary conditions for high-fidelity single-atom preparation and readout, laying the groundwork for the implementation of advanced programmable Rydberg-based quantum simulators with alkaline-earth Sr neutral atoms.

\vspace{6pt} 

\authorcontributions{Conceptualization and methodology, V.G. (Vladislav Gavryusev), J.C., G.C. and L.F.; software, A.F.; validation, V.G. (Veronica Giardini), A.F. and V.G. (Vladislav Gavryusev); formal analysis, A.F. and V.G. (Veronica Giardini); investigation, V.G. (Veronica Giardini), L.G., A.F. and S.S.; data curation and visualization, V.G. (Veronica Giardini) and A.F.; writing---original draft preparation, V.G. (Veronica Giardini), A.F., L.G. and V.G. (Vladislav Gavryusev); writing---review and editing, all authors; supervision, project administration, resources and funding acquisition, V.G. (Vladislav Gavryusev), M.I., J.C., G.C. and L.F.. All authors have read and agreed to the published version of the manuscript.}

\funding{This project has received funding from Consiglio Nazionale delle Ricerche (CNR) PASQUA Infrastructure, QuantERA ERA-NET Cofund in Quantum Technologies project MENTA, from the Italian Ministry of Education and Research (MUR) PRIN 2022SJCKAH "HIGHEST", and, in the context of the National Recovery and Resilience Plan and Next Generation EU, from M4C2 investment 1.2 project MicroSpinEnergy (Vladislav Gavryusev).}

\dataavailability{The original contributions presented in this study are included in the article. The raw data supporting the conclusions of this article will be made available by the authors on request. Further inquiries can be directed to the corresponding author.} 

\acknowledgments{We acknowledge insightful discussions and collaboration on the initial planning steps with Francesco Scazza. We thank the mechanical and electronic workshops of LENS and of the Physics and Astronomy Department for their support.}

\conflictsofinterest{The authors declare no conflict of interest.}

\isPreprints{}{
\begin{adjustwidth}{-\extralength}{0cm}
} 

~\reftitle{References}

\bibliography{PaperExpSet}

\PublishersNote{}
\isPreprints{}{
\end{adjustwidth}
} 
\end{document}